# THE RIEMANN MAGNETON OF THE PRIMES

Danilo Merlini

Cerfim[1] and Issi[2]

Via F.Rusca 1, P.O.Box 1132
6600 Locarno, Switzerland
merlini@cerfim.ch

**Abstract**

The aim of this paper is to present some aspects of the complexity related to the Riemann Hypothesis (RH). We describe a calculation which involves a function related to the Riemann Zeta function and suggested by two recent papers concerning the Riemann Hypothesis: one by Balazard, Saias and Yor (1999) and the other by Volchkov (1995). We define an integral $\varphi(\rho)$ involving the Zeta function in the complex variable $s = \rho + i \cdot t$ and find a particularly interesting expression for $\varphi(\rho)$ which is rigorous at least for some range of $\rho$. In such a range we find that there are two discontinuities of the derivative $\varphi'(\rho)$ at $\rho = 1$ and $\rho = 0$, which we calculate. The jump at $\rho = 1$ is given by $4\cdot\pi$, while that at $\rho = 0$ is given by $\pi\cdot(-4 +\gamma + 3\cdot\ln(2) +\pi/2)$. The validity of the expression for $\varphi(\rho)$ up to $\rho = \frac{1}{2}$ is equivalent to the truth of the Riemann Hypothesis. Assuming RH, the expression for $\varphi(\rho)$ gives $\varphi = 0$ at $\rho = \frac{1}{2}$ and the slope $\varphi'(\rho) = \pi ( 1 + \gamma) = 4.95..$ at $\rho = \frac{1}{2}$ (where $\gamma = 0.577215...$ is the Euler constant). As a consequence, if the expression for $\varphi(\rho)$ can be continued up to $\rho=\frac{1}{2}$, if we interpret $\varphi(\rho)$ as the effective potential of a charged filament given by $f = \rho+ i \cdot t$, ($t\in R$, $\rho$ constant), in the presence of the unit charges located at all the zeros (trivial and non trivial zeros) and at the pole in 1 of the Riemann Zeta function, then there is a jump of the electric field $E (\rho) = \varphi'(\rho)$ at $\rho = \frac{1}{2}$ given by $2\cdot\pi$ times the first Li's coefficient $\lambda_1 = (1+\gamma/2-(1/2)\ln(4\cdot\pi)) = 0.0230957…$ We then construct a potential well ( a symmetric function around $x = \rho+\frac{1}{2} = 1$ ) which is exact if the RH is true. Independently of the RH, by looking at the behaviour of the convergent Taylor expansion of $\varphi(\rho)$ at $\rho = 1-$, the value $\varphi(\rho = \frac{1}{2}+)$ as well as the first Li's coefficient may be evaluated using the Euler product formula. We give in this way further evidence for the possible truth of the Riemann Hypothesis.

[1] Research Center for Mathematics and Physics
[2] Institute for Scientific and Interdisciplinary Studies

# 1. Introduction

In recent years there has been a growing interest in the connection between physics and arithmetics and much effort has been devoted to obtaining some new light concerning the underlying phenomena, which should explain some very basic and fundamental properties of the natural numbers like the primes. In this case study, many areas of mathematics and physics are involved..
The properties of the distribution of the prime numbers are encoded in the Riemann Zeta Function $\zeta(s)$ of the complex variable $s = \rho + i \cdot t$ in the plane, $\rho, t \in R$ which we look at as a "partition function" for a certain model of statistical mechanics.. In this sense, for $\rho > 1$, the "Riemann partition function" $\zeta(s)$ is given by the Euler product of "small partition functions" $1/(1-(1/p)^{(\rho+i.t)})$ which involve the primes ($p = 2, 3 \ldots$)

$$\prod_p 1/(1 - 1/p^{(\rho + i \cdot t)})$$

$\zeta(s)$ has no zeros in such a domain. For $\rho < 0$, $\zeta(s)$ has an infinite set of real zeros, called the trivial zeros, which are the negative even integers, $s = \rho = -2 \cdot n$, $n \in N^*$.
The other zeros constitute the "non trivial set", and are known to lie in the critical strip $|\rho - ½| < ½$; the Riemann Hypothesis says that these are all on the line $\rho = ½$. Many fundamental works of these recent years have given some evidence that the non trivial zeros are related to and should constitute the spectrum, as quantum levels or energies, of an unknown dynamical physical system. The statistical properties, like the correlation functions of the non trivial zeros of $\zeta$, are strongly connected with the fluctuations of the distribution of the primes governed by the prime counting function $\pi(x)$ which gives the number of primes less or equal to x, for any real number x We mention here some (among others ) very important works of a different nature, all of which explain the complexity related to the Riemann Hypothesis, as well as the recent advances in the field at a high mathematical level : see Riemann (1900), Zagier (1981), Odlyzko (1987,1990 and 2000), Montgomery(1973), Alexander, Baglawski and Rota(1993), Bombieri (2000),Connes (1999),Cohen (1998); see also Berry and Keating (1998, 1999) for a very stimulating heuristic work. In particular, a fundamental connection of the RH with statistical mechanics of classical spin systems has been found and investigated recently, see Knauf (1993), Contucci and Knauf (1997). In fact, it has been possible to construct a one dimensional spin system with many body interactions, the partition function of which, in the complex temperature s, is the ratio of two Riemann partition functions, and many interesting rigorous results have been obtained. Also, of special interest in the context of statistical mechanics is the result, Knauf (1993), that the many body interaction in the one dimensional spin chain are of the ferromagnetic type and become translation invariant in the corresponding thermodynamic limit.
The purpose of this work is to present some analytical calculations followed ,for illustration, by numerical experiments carried out on a function which, in the spirit of classical mechanics may be seen as a mean potential energy of a vertical charged filament in the plane (placed at $\rho$) which we are free to call the "potential" of a natural "model" suggested by the versions of the two independent system formulations of the RH (the second of which is related to the Brownian motion process, see Volchkov(1995), Balazard, Saias and Yor (1999). We will call such a "model", given in terms of a certain function φ , the Riemann Magneton of the Primes . Since analytic and numerical calculations relating to the theorem by Balazard et. al (1999) and the theorem of Volchkov (1995) have, to the best of our knowledge not appeared in literature, we thought it might be of some interest to present them. The idea of getting some information using the Taylor expansion for a function related to the Riemann Zeta function ( here the function φ(ρ) ) related to the Riemann Zeta function ζ inside the critical strip was also advanced in a recent work, which was mainly concerned with a calculation of the Riemann-Mangoldt function using relatively high values of the primes, Merlini and Rusconi (2002,2004) .In the simple calculations we present, the restriction ρ = ½ is avoided ( Balazard, Saias and Yor(1999) ).

If we extend such an expression inside the critical strip we find that the potential φ(ρ) is continuous as a function of ρ but two singularities are present, at ρ = 0 and at ρ = 1, and these may be interpreted as jumps of the electric field in the interpretation that φ is the potential energy of the charged filament .Moreover, if we assume that the Riemann Hypothesis is true, then there is another jump at ρ = ½ . The first two jumps are a consequence of the application of the prime number theorem to our computations while the second is a consequence of the reflection of the Brownian motion in the representation given by Balazard,Saias and Yor (1999) ,which implies on computations a shift of ½ along the real axis.  Even if we  do not assume the RH to be true ,our rigorous solution at the border of the critical strip has nevertheless a convergent Taylor expansion with radius of convergence equal to ½ and this allows us to compute a constant (the first Li's coefficient $\lambda_1$ ) using the primes. The Taylor series of the real function φ(ρ),contains all Li's coefficients and the first, $\lambda_1$ ,has the meaning of being exactly the transversal electrical field filled by the test charged filament placed at ρ=½.This field is felt by an observer at the origin in the complex plane and is due to the "discrete charges" sitting at the location of the nontrivial zeroes (non trivial charges), those sitting at the location of the trivial zeros and that located at the pole of the Riemann Zeta function ;the potential energy is given in terms of the two dimensional Coulomb logarithmic potential between two charges (here in the plane) and is similar to the one appearing in the one dimensional Mehta-Dyson model, see Mehta (1967), Merlini, Rusconi and Sala (1999) . The main result of this work is then given at the end of Section 3 in the form of a theorem .Our analysis gives a reformulation of the RH in terms of the Coulomb potential and in this way further evidence in the direction that the RH should be true.

## 2.The formulation and the computations

Volchkov (1995) has shown that the truth of the Riemann Hypothesis ( RH ) is equivalent to a computation in the complex plane of the integral of log(|ζ(s)|) (where s is the complex variable given by s = ρ +i·t ) , modulated with some weight μ(t ) = 1/(¼+t$^2$) ; we call μ(t)·dt the "Lorentz. measure". Balazard, Saias and Yor (1999) proposed a criterion that involves an integration of the Riemann Zeta function in the complex plane in one variable along ρ = ½ only For new computations techniques connected with discussions of the RH, see Odlyzko ( 1987, 1990 and 2000 ).Using these considerations we start our treatment to construct the function φ(ρ). A recent result in the form of a theorem, which should simplify the problem of the computations and calculus with respect to the equally rich criterion for the RH given in Volchkov (1995) has been obtained by Balazard,Saias and Yor (1999).The theorem states that the truth of the RH ( the Riemann Hypothesis) is equivalent to following condition :

$$\varphi(1/2) = (1/2) \cdot \int_{\rho=1/2} dt \cdot \log(|\zeta(s)|) \cdot (1/(1/4+t^2)) = 0 \qquad (1)$$

The above result is based on advanced mathematics and is connected with Brownian motion and Hardy Spaces as explained by the authors. Eq.(1) contains an integration only in one variable, i.e. on the critical line. The connection with the criterion given in Volchkov (1995) will be discussed in the light of the function φ(ρ), that we introduce in this work. Both the results mentioned above are interesting and they inspired the present work, which is concerned with simple mathematics but which allows us, at the end ,to make some connection with other additional recent rigorous works, the results of which lead us to believe that the RH should be true, see Pustyl'nikov (1999), Li (1997), Bombieri and Lagarias (1999), Lagarias (1998).

Eq.(1) may be interpreted as the effective "potential" of a charged filament placed on the critical line ρ = ½ .We now introduce a perturbation on φ (½) with respect to the critical line , keeping μ(t).dt as the "partition measure". This is achieved simply by moving the charged filament to the right or to the left of the critical line, i.e. at any position ρ and in the sequel we investigate, by simple analytical computations, the potential φ(ρ) (for ρ>½).Some numerical computations are presented for illustration. We consider $s = \rho + i \cdot t$ instead of $s_0 = ½ + i \cdot t$ because we are looking at the difference $s - s_0 = \rho - ½$ as a displacement of the charged filament The potential φ(ρ) is then expected to be positive at positions ρ>½ and negative for ρ<½. If there is no jump of φ at ρ = ½ then the potential φ is expected to vanish according to the Riemann Hypothesis as stated by Eq.(1) This does not mean that its derivative, the electric field, is continuous. Below, we first show that the potential will be continuous at ρ= ½ if one assume that φ(½) = (½)· (φ(½+)+φ(½-)) i.e. in particular at ρ = ½, as a consequence of the "duality relation" i.e. the symmetry relation s ->1-s, which is a well known property of the Riemann Zeta function (Riemann symmetry ).Even if φ(ρ) were continuous at ρ=½. ,the question is still open whether φ is differentiable at ρ=½ .For any s ∈ C, the function ξ(s) is given by the following expression (Patterson , 1995; Titchmarsh 1999)

$$\pi^{(-1/2 \cdot s)} \cdot s \cdot (s-1) \cdot \Gamma(s/2) \cdot \zeta(s) = \xi(s) \qquad (2)$$

and $\qquad \xi(s) = \xi(1-s) \qquad (3)$

( where ξ is the Riemann's Xi function, see e.g. Titchmarsh ( 1999 )).

The above relation is valid in particular for values of $\Delta = \rho-½$ such that $|\Delta| = |\rho-½| < ½$ . (critical strip). For any ρ the function φ is given by

$$\varphi(\rho) = (1/2) \cdot \int_R dt \cdot \ln(|\zeta(s)|) \cdot (1/(1/4 + t^2)) \qquad (4)$$

Looking at Eq.(2), written explicitly in the representation with the nontrivial zeros ($s_0$) and the trivial zeros ($r_0$) ( the poles of the Gamma function) , we have that (Patterson, 1995; Titchmarsh, 1999):

$$\varphi(\rho) = (1/2) \cdot \int d\mu(t) \cdot \ln(|\zeta(s)|) =$$

$$= (1/2) \cdot \int d\mu(t) \cdot \ln\left|\prod (1 - s/s_0) \cdot (1 - s/(1-s_0)) \cdot \prod (1 - s/r_0) \cdot \exp(s/r_0) \cdot (1/(s-1)) \cdot (\pi^{s/2} \cdot e^{\gamma \cdot s/2})\right|$$

so that a part of a linear potential in ρ involving γ and ln(π) in the above Formula, there is the appearance of the logarithmic 2d Coulomb potential with a plus sign for each zero and with a minus sign for the pole of the Riemann Zeta function located at s = 1.Integration along the charged filament line, with the Lorentz weight µ(t) gives then rise to an effective potential for a test particle sitting on the real axis shifted of exactly ½ to the right if ρ > max (1,Re($s_o$), Re($r_o$) ). In the above products $s_o$ is any non trivial zero and $r_o$ any trivial zero of ζ(s).Thus φ(ρ) is the potential energy of a test particle sitting on the real axis in the world of the charges (zero's and pole) distribution of the Riemann Zeta function.

We now use the symmetry of the function ξ(s) to obtain a relation for the potential which holds independent of the truth of the RH. In fact, since  ξ(s) = ξ(1-s) , then | ξ(ρ +i· t) | = |ξ(1-ρ- i · t)| = |ξ(1-ρ+i.t)|.

Above we used the Formula for the Gamma function given by

$$1/\Gamma(s) = s \cdot \exp(\gamma \cdot s) \cdot \prod_{n=1}^{\infty}[(1+s/n) \cdot \exp(-s/n)]$$

where γ is the Euler constant .We may also use the following formula (Gradshteyn and Ryzhik , 1965)

$$\int_0^\infty dt \cdot \ln(\beta^2 + \mu \cdot t^2) \cdot (1/(\alpha + t^2)) = \pi \cdot [1/(\alpha)^{1/2}] \cdot \ln[(\mu \cdot \alpha)^{1/2} + \beta], \beta \geq 0.$$

In particular, if β= ½, µ = 1 and α = ¼, the integral vanishes. Introducing the function f(ρ) = φ(ρ) – φ(1-ρ ), we obtain the symmetry relation for φ given by:

$$f(\rho) = \varphi(\rho) - \varphi(1-\rho) = \pi \cdot (\ln(\pi) \cdot (\rho - ½) + \ln(\Gamma(¼ + ½ \cdot |\rho - 1|)/\Gamma(¼ + ½ \cdot |\rho|))) \quad (6)$$

The above relation shows, in particular, that φ(ρ) is continuous at ρ=½ since the right hand side of this equation vanishes as ρ↓½ . Even if φ(ρ) is continuous at ρ=½, the question is still open whether φ is differentiable at ρ=½.. In fact if φ(ρ) were differentiable at ρ = ½ , then ,using Eq.(6),the slope at that point would be given by :

φ'(½) = (½)·f '(½) = (½)·π·(lnπ - (Γ'/Γ)(½)) = π·(ln π +γ + 2.ln2)/2 = 4.882410

(in fact, assuming the RH, we know that φ is not differentiable at ρ = ½ ( see Eq.(14) below). Now, if φ(1-ρ) = 0 , then φ(ρ) = f(ρ) .We note that the numerical experiments for φ(ρ), given below, indicate that φ( ρ ) is very different from f(ρ) except in a small interval around ρ = ½ where φ(ρ) could vanish. Moreover, around ρ = ½+, φ (ρ ) appears to be increasing .

The situation is illustrated by means of some numerical computations.
Table 1 contains the computations of φ(ρ) in a large range ( numerically) where the t values run from 0 to no more than 30-50.

| ρ | φ | | | | |
|---|---|---|---|---|---|
| 1 | 3.01779 | | | | |
| 0 | -2.23087 | φ(1) – φ(0) | = 5.24865 | exact f(1) | = 5.20595 |
| 0.8 | 1.63917 | | | | |
| 0.2 | -1.37146 | φ(0.8) – φ(0.2) | = 3.01064 | exact f(0.8) | = 2.99220 |
| 0.7 | 1.05207 | | | | |
| 0.3 | -0.93122 | φ(0.7) – φ(0.3) | = 1.98329 | exact f(0.7) | = 1.97100 |
| 0.6 | 0.50914 | | | | |

| | | | |
|---|---|---|---|
| 0.4 | -0.48939 | φ(0.6) – φ(0.4)  = 0.99853 | exact  f(0.6) = 0.97869 |
| 0.55 | 0.25132 | | |
| 0.45 | -0.24100 | φ(0.55) – φ(0.45) = 0.49233 | exact  f(0.55) = 0.48851 |
| 0.5 | 0.00026 | | |
| 1.2 | 2.26182 | exact  φ(1.2) = | 2.261725 |
| 1.5 | 1.56362 | exact  φ(1.5) = | 1.563571 |
| 2 | 0.92295 | exact  φ( 2 ) = | 0.922933 |
| 2.5 | 0.57815 | exact  φ(2.5) = | 0.578160 |
| 3 | 0.37485 | exact  φ( 3 ) = | 0.374864 |
| 4 | 0.16732 | exact  φ ( 4) = | 0.167332 |
| 6 | 0.03748 | exact  φ ( 6 ) = | 0.037493 |

Table 1 Numerical experiments

The above numerical experiments turn out to be controlled by the symmetry relation given  by   the function f(ρ) and are given for illustration. Outside the right border of the critical strip the numerical experiments also agree well with the exact results given by the exact expression  obtained below (se Eq (7) and  Eq.(8)), as the next  step.

The first observation is that, since φ(1) > 0 and φ(0) < 0,  φ(ρ) should vanish at least  for one value of ρ, if φ is continuous. This is supported by the numerical experiments. Second, we note that φ is anti-symmetric around ρ = 0.5 , a manifestation that  φ might indeed vanish  for ρ = 0.5, but a proof of this is still lacking, since the RH has not yet been proven. (These numerical computations, of course, have another goal and have nothing to do with those  carried out  in the critical strip for many millions of zeroes of the Zeta function ,which constitute   advanced   numerical research in the field (Odlyzko, 1990)) .

### 3.Analytical computations

An  exact expression for φ(ρ) outside the critical strip ( for ρ > 1 and thus by the symmetry relation also for ρ < 0 ) may   easily be obtained   using the Euler  product. In fact , for ρ > 1 one has

$$\varphi(\rho) = (1/2) \cdot \int_R dt \cdot \ln(|\prod 1/(1 - 1/p^s)|) \cdot (1/(1/4 + t^2))$$

where R is the real line.

In the complex  t- plane ,the poles are located at  t = ½·i and at t = -½·i,  and the integral in the complex plane of t may be carried out ; a shift appears along the real axis ρ ,of  exactly  ½  and we obtain

$$\varphi( \rho ) = \pi \cdot \ln(\zeta(\rho + \tfrac{1}{2})) \quad \text{for } \rho \geq 1 \tag{7}$$

Using the exact symmetry relation  we  find that the exact solution for φ(ρ) in the range   ρ ≤ 0 is given by

$$\varphi(\rho) = \pi \cdot \ln(\pi^{(\rho-\tfrac{1}{2})} \zeta(3/2 - \rho) \cdot (\Gamma(3/4 - \tfrac{1}{2}\rho)/\Gamma(1/4 - \tfrac{1}{2}\rho))), \quad \rho \leq 0. \tag{8}$$

(7) and (8) give  the potential energy for a test charge  (integral of the charged filament with density μ(t) ) located at ρ) thus located at   ρ+½  on the real axis.

For ρ > 1, φ is a decreasing function to o as ρ->∞ ,starting from φ(1) = 3.016.. For ρ < 0, φ is a negative increasing function of ρ which is almost linear, reaching the value φ(0) = -2.19..at ρ = 0. The exact values of φ outside the critical strip obtained above are unique and also conform to the experimental results; these are, of course, obtained by integrating ln (|ζ| ) which contains the secrets of the zeros, but as we have seen , the range for t in our computations varies from 0 to about 50. Moreover, the measure μ(t)·dt appearing in both works, Volchkov (1995), Balazard, Saias and Yor (1999), is responsible for giving fast convergent values of φ very close to the "thermodynamic limit"(t∈R) in the sense given by the exact solution given above. In Table 1 we have also given the experimental results for |ρ-½| < ½ .We now go into the critical strip and obtain an expression for φ(ρ) which explains here the above numerical results i.e. we obtain an expression for φ(ρ) , which is continuous and which necessarily matches the outside solution at the two borderlines ρ = 0 and ρ = 1 .The above mentioned expression for φ follows from simple analytical computations which, to the best of our knowledge, have not yet been given along the same lines .The shift of ½ is new and has not appeared in other computations on the zeroes of the Riemann Zeta function.

Using the definition of φ , given by (4) , we have :

$$\varphi(\rho) = (1/2) \cdot \int_R dt \cdot \ln \left| \pi^{(s/2)} \cdot (1/(s \cdot (s-1))) \cdot (1/\Gamma(s/2)) \cdot \zeta(s) \right| \cdot (1/(1/4 + t^2))$$

The first integral gives, using the formulas given above for the Gamma function , the contribution

$$(½)\cdot\pi \cdot \ln(\pi)\cdot \rho - \pi \cdot \ln(½+ \rho)\cdot(½+ |1-\rho|) + \pi \cdot \ln(1/\Gamma(¼ +½\cdot|\rho|))$$

For the second integral we use the formula | ξ ( s )| = ∏ |(1-s /(z))·(1- s/(1-z))| where the product extends to all non trivial zeros z = $z_0$ +i·$t_0$ in the complex upper half plane (Im(z) > 0), see Patterson (1995), and Titchmarsh (1999).

By integration of the above expression in the whole complex upper half plane of t, ( taking into account that there is a pole at t =i/2 and another one at t = $t_0$ + i·(ρ – $z_0$) ) ,we obtain

$$\pi \cdot \ln(1-(\rho+1/2)/z)\cdot(1-(\rho+1/2)/(1-z)) + g$$

where g is a function which is zero if ρ > $z_0$ . The presence of the shift of ½ allows us to re-write the Zeta function shifted by ½ (with the presence of two factors) so that

$$\varphi(\rho) = \pi \cdot \ln[\zeta(\rho+½))\cdot(\rho-½)/(|\rho-1| + ½)] + \Phi(\rho,\{z\}) \qquad (9)$$

where Φ = 0 if ρ > u = max{$z_0$}, with z = $z_0$ +i·$t_0$ any non trivial zero of ζ(s). In particular, if ρ> 1, we recover the expression given by Eq.(7) .

There is in Eq.(9) a well defined change ( as compared with Eq.(7) and Eq.(8) ) of the potential inside the critical strip given by ln[(ρ-½)/(3/2 – ρ )] and , of course, the presence of the unknown function Φ .The result given by formula (9) is rigorous, due to the Prime number theorem (see Titchmarsh (1999), Patterson (1995)), in an open set of the real axis strictly inside the critical segment [0,1] .
If we assume that Φ in Eq.(9) is strictly zero not only around ρ= 1⁻ but up to ρ= ½⁺ and assume the validity of the RH we find that

$$\varphi(\rho) = \pi \cdot \ln(\zeta(\rho +½)\cdot(\rho-½)/(|\rho-1| + ½)) \quad \text{for } \rho > ½ \qquad (10)$$

φ(1-ρ) = φ(ρ)– f(ρ) = π·ln(ζ(ρ+½)·(ρ-½)/(|ρ-1|+½))-π·ln(π)·(ρ-½)-π·ln(Γ(½|ρ-1|+¼)/Γ(½|ρ|+¼))
for 1-ρ = ρ' < ½ , that is

$$\varphi(\rho) = \pi \cdot \ln[\text{Zeta}(3/2-\rho)\cdot(1/2-\rho)/(1/2+|\rho|)] - \pi \cdot [\ln(\pi)\cdot(½-\rho) + \ln(\Gamma(¼+½|\rho|)/\Gamma(¼+½|1-\rho|))] \quad (11)$$

for ρ < ½ . Before presenting the plot of φ, we give the values of the potential φ calculated with the expression (9) and (10) above in Table 2 .The values agree well with the results of the numerical experiments given above in Table 1.

| φ    | ρ         |
|------|-----------|
| 1    | 3.016745  |
| 0    | -2.189208 |
| 0.8  | 1.639400  |
| 0.2  | -1.396331 |
| 0.7  | 1.052322  |
| 0.3  | -0.918670 |
| 0.6  | 0.509444  |
| 0.4  | -0.4692   |
| 0.55 | 0.251081  |
| 0.45 | -0.237435 |

Table 2  The potential φ(ρ) in the critical strip (Eq(10) and Eq.(11)).

In Fig.1 we give the plot of φ(ρ) as a function of ρ in the whole range of ρ.(Eq.(10) and Eq.(11). Observing Fig.1 we note that for negative values of ρ , φ is negative, mostly linear (due to the presence of the uniform spectrum of the trivial zeros of the Riemann Zeta function for ρ<0),increasing to the values φ(0) = -2.1892.., to ρ = 0 ; here we encounter the chaotic spectrum (discussed by other methods e.g. in ( Berry and Keating (1999)) ) of the zeros of the Zeta function in the critical strip, φ is concave ,vanishing at ρ = ½ , and reaching the value φ = 3.016 at ρ= 1. At ρ > 1 the he Zeta function is given by the Euler product and φ(ρ) decreases monotonically to zero as ρ → +∞.

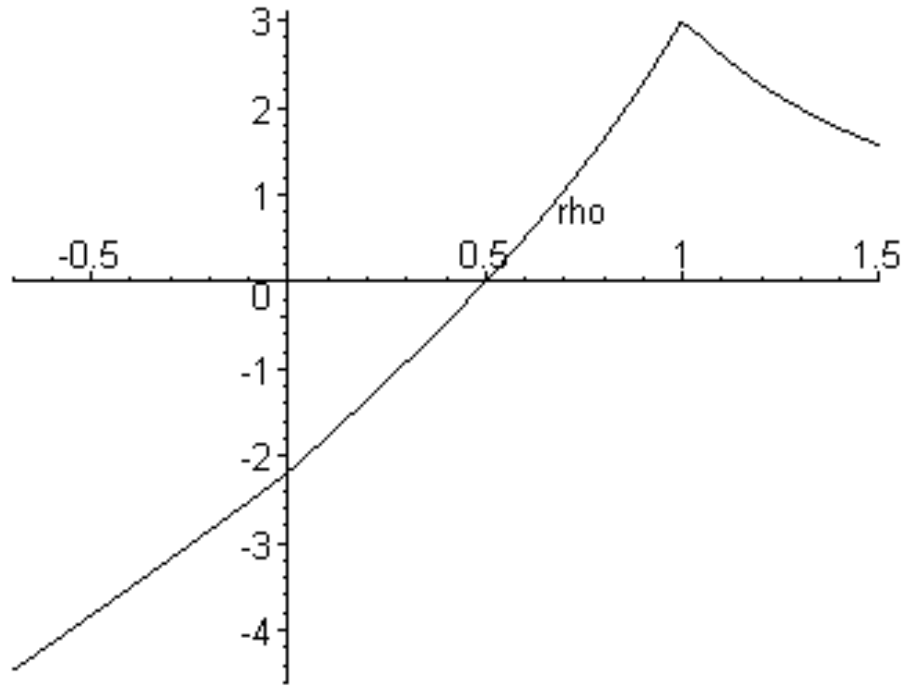

Fig.1 Plot of the potential φ(ρ) (Eq.(10) and Eq.(11) where x = ρ and y = φ(ρ)).

From Eq.(10) it follows that the potential φ(ρ) vanishes exactly at ρ = ½, since

$$\pi \cdot \ln\ [\zeta(\rho+½)\cdot(\rho-½)/(3/2-\rho)]\ =\ \pi \cdot \ln\ [(\zeta(x)\cdot(x-1)] = 0, \quad \text{as } x = \rho+½ = 1^{+}$$

so that the vanishing of φ(ρ) at ρ = ½ is ensured here by the behaviour of the Riemann Zeta function at its unique pole ρ = 1 +0·i ( of course the above formula may also be written using the ξ function, see Fig.5). As anticipated above, it should be noted that (assuming the RH) φ is not differentiable at ρ = ½ (the jump of the electrical field given by E = φ'(ρ) at ρ=½ is however invisible on the plot of Fig(1)) ; φ(ρ) is also not differentiable at ρ = 1 and at ρ = 0. On Fig. 2 below we present the plot of the transversal electrical field of the magneton given by E(ρ)=φ'(ρ).

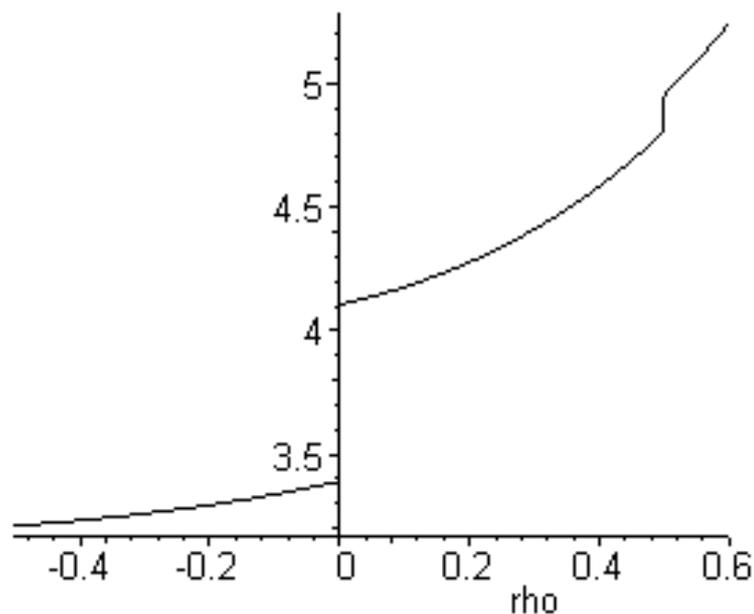

Fig.2 Plot of the transversal electrical field E(ρ) as a function of ρ.

(following Eq.(7)-Eq.(8),Eq.(10) and Eq.(11)),up to ρ = 0.6;
jumps at ρ = 0 and at the critical line ρ = ½).

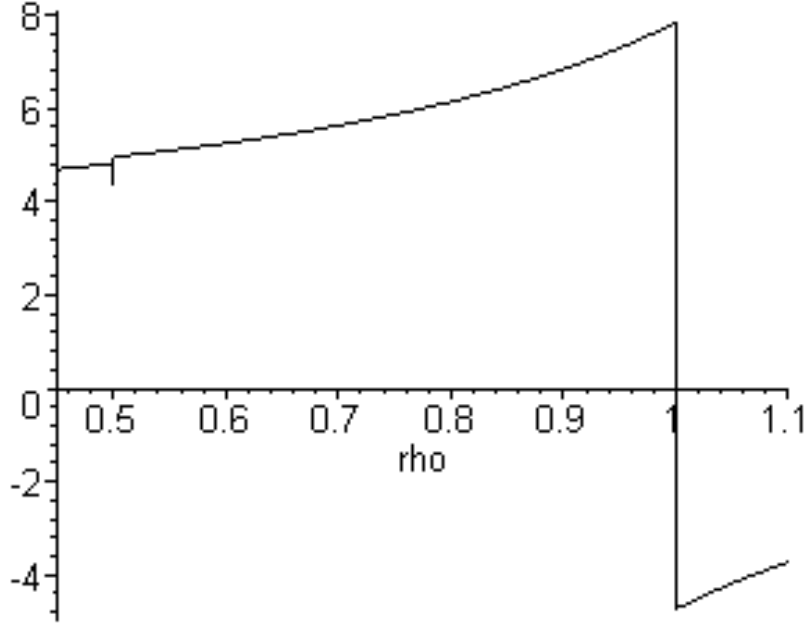

Fig.3. Plot of the transversal electrical field E(ρ) as a function of ρ
from the critical line up to ρ = 1.1 (jump of 4π at ρ=1)

Formulas (10) and (11) are only correct in the whole range of ρ if we assume that the RH is true. However they hold even without assuming the RH, for an open set strictly inside the two borderlines of the critical segment ( ρ=0$^+$ and ρ = 1$^-$) (this uses as mentioned above the result of Balazard, Saias and Yor (1999) and the Prime Number Theorem) and as discussed below, formulas (10) and (11) are exact up to $10^{-20}$.

To give a more symmetric "electrostatic picture" of Eq.(10) in terms of the trivial and non trivial zero, and the pole of the Riemann Zeta function using the 2-d Coulomb potential , we construct the symmetric function S(φ(ρ)) = (1/2)·( φ(ρ)+φ(1-ρ)) ; then using Eq.(10) and Eq.(6), and using the formula for the product of two Gamma functions given in Zagier (1981) that is

$$1/(\Gamma(z)\cdot\Gamma(1-z)) = \sin(\pi\cdot z)/\pi = (\sin(\pi\cdot z)/(\pi\cdot z))\cdot z = z\cdot\prod_{n=1}^{\infty}(1 - z^2/n^2)$$

where z = (ρ+½)/2 and 1-z = (3/2-ρ)/2 , we obtain:

S(ρ) = (π/2) ln((ξ(ρ+½))²·(π/2)[(ρ+½)·(3/2-ρ)]$^{-3/2}$ · (Π (1-(ρ+½)²/4·n²)·Π(1-(3/2-ρ)²/4n²))$^{1/2}$

where the product runs from n=1 to n=∞ ,so that with the variable change x=ρ+½, , we have that

S(x) = π· (∑ln│(1-x/s$_o$)·(1-x/(1-s$_o$))│+(1/4)·(∑ln│(1-(x/2n)²)(1-[(2-x)/2n]²)│-(3/4)·ln(x·(2-x)+(1/2)ln(π) )

where s$_o$ is any nontrivial zero of the Zeta function and n is any positive integer.

This shows the perfect balance between the charges (zeros) for a test charge situated at x and its image situated at the symmetric point 2-x where the symmetry axis is at x = 1. For x=1, since the above formula is based on Eq.(10) (valid up to x=ρ+½ = 1, assuming the RH), we have S(x=1) = 0 as for φ ( at x=1 the potential with the charges sitting at the place of the trivial zeros vanish, the same that due to the pole and its symmetric at 2-x (negative charges) and that due to the charges located on the critical line, assuming that the RH is true). The plot of the potential well S(x) is given in the Fig 4.

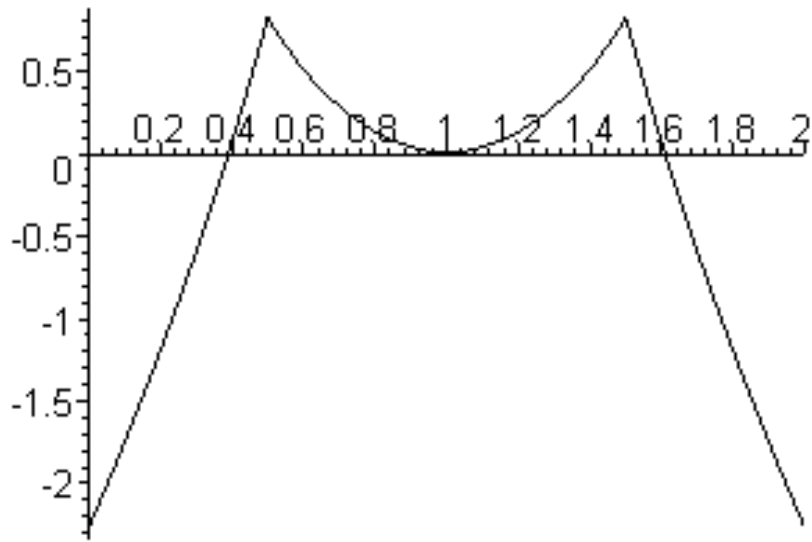

Fig.4 Plot of the function S(x) symmetric around x = 1.

Outside the critical interval x=]0,1[, there are exactly two solutions of the Equation S(x)= 0, given by the solutions of the Equation:

$$\zeta(x)^2 \cdot (\Gamma(x/2)/\Gamma((x-1)/2)) \cdot \pi^{(1-x)} = 1$$

Thus, $x_1 = 1.610217484$ and $2-x_1 = x_2 = 0.3897825164$. The mean value is x=1 and S(1) = 0 is equivalent to the RH (see on Fig 4 the bottom of the potential well). For the two values $\rho_1 = x_1 - \frac{1}{2} = 1.11021\ldots$ and $\rho_2 = x_2 - \frac{1}{2} = -0.11021\ldots$; one verifies on Fig.1 that $\varphi(\rho_1) = -\varphi(\rho_2)$ and that the straight line between the two points of the curve passes through the midpoint of the critical segment given by $\rho = \frac{1}{2}$. (the latter without assuming the RH !).
As illustration we may now use $x_1$ in the above Equation to obtain within some decimals that

$$|\xi(x_1)| = \Pi (1-x_1/s_o) \cdot (1-x_1/(1-s_o)) = 1.022934630..$$

where the product is taken as before, over all nontrivial zeros with $Im(s_o) > 0$.

Another criterion for the RH is given by the theorem of Volchkov, and we now make the connection recalling first the theorem-criterion of Volchkov (1995). It was proven by Volchkov that the Riemann Hypothesis is equivalent to the assertion that

$$\int d\rho \int dt \cdot \ln|\zeta(\rho + i \cdot t)| \cdot (1 - 12 \cdot t^2) \cdot (1/(1 + 4 \cdot t^2))^3 = \pi \cdot (3 - \gamma)/32 \quad (12)$$

where $\gamma = 0.577215\ldots$ is the Euler-Mascheroni constant.
In Eq.(12), the integral extends in the half plane right from the border of the critical line, i.e $\rho > 1/2$
Using the function $\varphi$ as defined by Eq.(4), it may be shown that such an equivalence can be written, using $\varphi'(\infty) = 0$, as

$$\delta = \varphi'(1-) - \varphi'(\tfrac{1}{2}+) - \varphi'(1+) = \pi(3 - \gamma) \quad (13)$$

As stated above, $\varphi$ is not differentiable at the two borders of the critical strip nor at $\rho = \tfrac{1}{2}$. Eq.(13) says that the total change of the slope (total jump of the electrical field E, between $\rho = \tfrac{1}{2}^+$ and $\rho = 1^+$) is given by the value of $\delta$. In fact our analytical solution (Eq.(10)) gives exactly:

$$\varphi'(1-) - \varphi'(1+) = \pi(\,(\zeta'/\zeta)(3/2+) - (\zeta'/\zeta)(3/2-)\,) = 4 \cdot \pi \quad (13')$$

The above value is independent of the truth of the RH. So, $\varphi(\rho)$ of Eq.(10) is not differentiable at $\rho = 1/2$ and Eq. (10) gives exactly $\varphi'(1/2+) = \pi \cdot (1+\gamma)$ where $\gamma$ is the Euler-Mascheroni constant. Thus:

$$\delta = \pi \cdot (4 - 1 - \gamma) = \pi \cdot (3 - \gamma)$$

in agreement with the value given by Volchkov.
The Taylor expansion of $\varphi(\rho)$ (given by (10)), has a convergent radius of $\tfrac{1}{2}$ and give fast convergent values to zero for $\varphi(\rho)$, as $\rho = \tfrac{1}{2}^+$. Here the Taylor expansion (assuming RH is true) is given by

$$\varphi(\rho) = \pi \cdot (\gamma + 1) \cdot (\rho - \tfrac{1}{2}) + O(\rho - \tfrac{1}{2})^2 \qquad \text{as } \rho = \tfrac{1}{2}^+.$$

and $E(\tfrac{1}{2}^+) = \varphi'(\tfrac{1}{2}^+) = \pi \cdot (\gamma + 1) = 4.954967\ldots$ so that our expression for $\varphi(\rho)$ is not differentiable at $\rho = \tfrac{1}{2}$ and the jump at the critical line is given (using Eq.(6) for $\varphi'(1/2)$) by

$$\Delta \varphi = \varphi(\tfrac{1}{2}^+) - \varphi(\tfrac{1}{2}^-) = 2 \cdot (\varphi(\tfrac{1}{2}^+) - \varphi(\tfrac{1}{2})) = 2 \cdot \pi \cdot (1 + \gamma/2 - \ln(4\pi)/2) = 2 \cdot \pi \cdot 0.0230957 \quad (14).$$

The above constant is given by the sum $\sum 1/[(s)(1-s)]$ over the nontrivial zeros belonging to Z (Patterson, 1995; Titchmarsh, 1999).
Here the meaning is that the potential of what we have called the "Riemann Magneton" of the Primes (assuming the RH is true), that is the expression (10) of the potential $\varphi(\rho)$ up to $\rho = \tfrac{1}{2}$, has a jump of its electrical field $E = \varphi'(\rho)$, at $\rho = \tfrac{1}{2}$, in addition to the two jumps we have rigorously proven at $\rho = 1$ and at $\rho = 0$ (related to the Prime number theorem and not to the truth of the RH.). Moreover, if at each nontrivial zero we "attach" a unit electric charge, then the constant of (14) (neglecting the factors 2 and $\pi$) is here interpreted, following the electrostatic picture given above, as the transversal electric field along the real axis at $\rho$, filled by a charged filament and this constant has been known for some years (Davenport, 1967; Volchkov, 1995; Bombieri and Lagarias, 1999; Patterson, 1995; Titchmarsh, 1999). It is now important to discuss our calculations in the light of very important works involving new rigorous mathematics and which go in the direction of a rigorous proof of the RH (Pustyl'nikov, 1999; Li, 1997; Bombieri and Lagarias, 1999).
It has been proven that the truth of the RH is equivalent to the proof that the Li's coefficients (Li, 1997; Bombieri and Lagarias, 1999), are all semi-positive and the explicit formulas have been derived (Bombieri and Lagarias, 1999). Of interest here is the first Li's coefficient, called $\lambda_1$. This coefficient turns out to coincide with the jump given above as can be verified by looking at Bombieri's and Lagarias's work. Our computation uses the rigorous solution for φ in an open set inside the right border of the strip (Eq.(10)). With the help of the function ξ, (the Xi function), using the variable $x= \rho +½$, (10), can be written (assuming RH is true), as

$$(1/\pi)\cdot\varphi(x-½) + \ln(\Gamma(x/2)\cdot(2-x)x\cdot\pi^{(-x/2)}) = <\ln|\xi(\rho+i.t)|>_\mu = \ln|\xi(x)|$$

and in terms of the ξ function (of argument x), the jump discussed above is located at x=1.
The above relation has been proven for a small open set around the value $x = 3/2$. So, without assuming the RH is true and without any other additional assumption all we can do is to analyse the Taylor expansion of $<\ln(|\xi(x)|)>$ around x=3/2, the aim being to obtain some information about $<\ln(|\xi(x)|)>$ around $x = 1$ as a function of the primes. The expansion was carried out to second order in the complex argument of ζ and it was found that the expansion gives satisfactory values at least for the low zeroes of ζ(s) (Merlini and Rusconi, 2004). Since the Taylor expansion of

$$<\ln(|\xi(x)|)> = <\ln(x\cdot(x-1)\cdot\Gamma(x/2)\cdot\zeta(x)\cdot\pi^{(-x/2)})> \qquad (15)$$

around x=3/2 exists, we have

$$<\ln(|\xi(x)|)> = \sum C_n\cdot(x-3/2)^n\cdot(1/n!), \qquad C_n = \partial^n<\ln(\xi(x))>/\partial x^n\ (x=3/2),$$

where the coefficients $C_n$ are well defined as functions of the primes.
Rearranging the well defined expansion around x=3/2, at x =1, we have the following

$$<\ln|\xi(x)|> = \sum C_n\cdot(1/n!)\cdot[\sum(x-1)^k\cdot(-½)^{n-k}\cdot B_{n,k}] \qquad (x=\rho+½)$$

where $B_{n,k}$ are the binomial coefficients.
Near $x = 1$ and up to $k = 2$, we then have the following asymptotic expansion:

$$<\ln(|\xi(x)|)> \sim \sum C_n(-½)^n\cdot(1/n!) + \sum C_n\cdot(-½)^{n-1}\cdot(n/n!)\cdot(x-1) + \sum C_n(-½)^{n-2}(1/2)\cdot n\cdot(n-1)(1/n!)\cdot(x-1)^2$$
$$(16)$$

and using some terms of the expansion (16), we obtain:

$$<\ln(|\xi(1)|)> \sim 0.8\cdot 10^{-5} \qquad \text{and}\ <(\xi'/\xi)(1)> \sim 0.023866$$

The second number is an approximation for the first Li's coefficient while the first indicates the possible vanishing of the potential φ at ρ= 0.5. With the first 13$^{teen}$ coefficients in the Taylor expansion we find that $<\ln(\xi(1))> = 0.157\cdot 10^{-19}$. So, if T is the value of the highest experimental zero which satisfies the RH, where T is roughly given by 2π·n/ln(n) for n = $10^{22}$ (Odlyzko, 1987), then one expects that m(½)<ln(T)/T, (using upper bounds of the type ln|ζ(s)|<c +ln.|s| for large values of |s|, Davenport (1967)), in Eq.(4)); then m(½) < 0.19 ·$10^{-17}$ and we note that the Taylor series for φ(1/2) or for $<\ln(|\xi(1)|)>$ give accurate values for the two number φ(½)= 0 and $\lambda_1$ = 0.0230957, (if the RH is true).
Let now suppose that all the zeros of the Zeta function satisfy the RH except one, given by z = a +i. ·To where ½ <a <1. Then in the integral for the corresponding factors in φ or in $<\ln|\xi|>$, we will have the amount ln(1- (ρ-½)/z) instead of the contribution ln(1-(ρ+½)/z), as for Eq.(10). In this case the total contribution to Φ of Eq.(9) will be given by Φ =2· ln[(1-(ρ-½)/z)/(1-(ρ+½)/z)]. For large values of $T_o$, we obtain (with x=ρ+½, at x=1+):

$$\Phi = 2 \cdot (a - \tfrac{1}{2})/T^2_o - 2 \cdot (x-1)/T^2_o \qquad (17)$$

This happens for each zero which does not satisfy the RH at first order in $(x-1)$ and in $1/T_o^2$. So the "slope" at $\rho = \tfrac{1}{2}$ or at $x = 1$ will be different from the one given by the first Li's coefficient or by that given by the Taylor series discussed above involving the primes. The correction will be very small, of the order of $-1.3 \cdot 10^{-42} \cdot (x-1)$ if $T_o$ is taken to be $1.25 \cdot 10^{21}$ corresponding to $n = 10^{22}$ as above. The value of the potential $\varphi(\rho)$ at $\rho = \tfrac{1}{2}$ (or of $< \ln(|\xi(x)|) >$ at $x = 1$) will be higher and one expects the slope to be smaller then that given by the first Li's coefficient..

We now assume that for $T > T_o$, all the zeros do not satisfy the RH ; then, the correction to the potential $\varphi(\rho)$ will be at most of the order $\ln(T_0)/T_0$ as shown below by direct computation. In fact, using Eq.(17), we have that:

$$\sum_{T \succ T_0}^{\infty}(c/T^2) = \int_{T_0}^{\infty} dN(t) \cdot (c/T^2) = \int_{T_0}^{\infty} dt \cdot (1/2\pi) \cdot (\ln T / T^2) = (c/2 \cdot \pi) \cdot (\ln T_0 / T_0) \qquad (18)$$

where we have used the well known result that $N(T) = (T/2\pi) \cdot \ln(T/2\pi)$ for large values of T (Patterson; 1995; Titchmarsh, 1999). So, the potentials $\varphi$ (the Riemann magneton of the primes), given by Eq.(10) and Eq.(11) as well as S (the potential well) are exact up to about $10^{-20}$, even in the whole critical segment [0,1], taking into account the recent advanced numerical experiments (Odlyzko, 2000 ).

Fig (5) shows the plot of the functions ( Eq.(15) and Eq.(16) ) where the tangent straight line has a slope given by the first Li's coefficient. The approximation given by Eq.(16) is of course invisible on the plot of Fig.5 even around $x=3/2$, since the function given by Eq.(15) is well described by two pieces of a parabola, thus illustrating the singularity of the first kind.

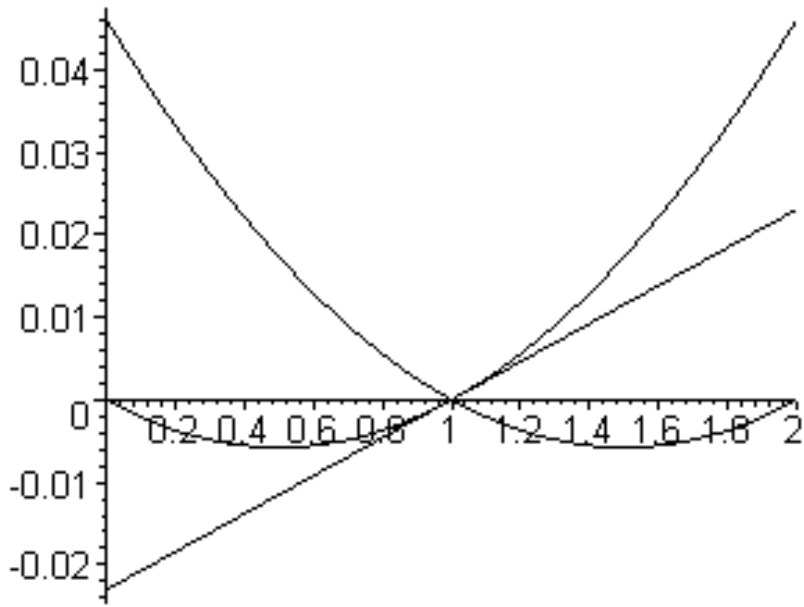

Fig.5 The plot of the function given by Eq.(15), symmetric around $x=1$

At this point we may subsume the main result of our analysis in the following theorem

**Theorem**

$$\varphi(\rho) \text{ be defined by } \varphi(\rho) = \tfrac{1}{2} \int_R dt \cdot \ln(|\zeta(s)|) \cdot 1/(1/4 + t^2)$$

We then have:

1. For any $\rho$ belonging to $]-\infty, 0+[ \cup ]1-, \infty[$, $\varphi(\rho)$ is given by Eq.(10) and Eq.(11). The derivative of $\varphi(\rho)$ that is the electrical field $E = \varphi'(\rho)$ has two jumps at $\rho = 0$ and at $\rho = 1$ which may be calculated exactly (See Eq.(13')).

2. If the expression given by Eq.(10) may be continued up to $\rho = \tfrac{1}{2}$, then Eq.(10) gives $\varphi(\tfrac{1}{2})=0$ and $\Delta E = \varphi'(\tfrac{1}{2}+) - \varphi'(\tfrac{1}{2}) = 2\cdot \pi \cdot \lambda_1$ (where $\lambda_1$ is the first Li's coefficient, $\varphi'(\tfrac{1}{2})$ is obtained by Eq.(6)) and the RH is true.

3. If for all $T \in [0, T_0]$, all the non trivial zeroes are on the critical line and above $T_0$ some or all non trivial zeroes are outside the critical line, we have:

$$\varphi(\tfrac{1}{2}) < c \cdot \ln(T_0)/T_0$$

where $c$ is some constant and $T_0$ is of the order $10^{21}$. Thus $\varphi(\tfrac{1}{2}+) < c' \cdot 10^{-20}$; the same holds for the symmetric potential well $S(x)$ of Fig.4, in particular at the bottom point given by $x=1$.

4. Independent of the truth of the RH, there are two positions for the test charge ie at $x_1$ and $x_2$ outside the critical segment, (whose centre is exactly given by $x = 1$, at the bottom of $S(x)$ on Fig 4), exactly where the symmetric function $S(x)$ vanishes.

**4. Concluding remarks and future work**

In this work we extended the definition of $\varphi(\tfrac{1}{2})$ of Balazard, Saias and Yor (1999) to any value of $\rho$ and named the new function the potential of *the Riemann Magneton of the Primes* (PRMP). (In our study the potential $\varphi$ may also be called a "free energy", because for $\rho > 1$, $\varphi$ has an interpretation in term of the spin chain model of Knauf (1993). With Eq.(10) and Eq.(11) we have constructed a function whose correctness is equivalent to the truth of the RH and we have shown that such a function is exact within the order of about $10^{-20}$ in the whole range of $\rho$. We have also given some connections with some recent works concerning the RH. Moreover an electrostatic picture involving the two-dimensional Coulomb potential has been given for the charged filament with density $d\mu$, with the appearance of a new potential well.
Concerning the Taylor expansion considered in Merlini and Rusconi (2004) which gives good approximations at least for the low nontrivial zeros (energy levels), it should be said that it is the change of slope at $\rho = 1^-$ (jump of the electrical field E) encountered in the present work that suggested the application of such an expansion for the potential. This is not visible and thus not suggested by the function (15) (plotted on Fig(4)) where the critical strip is invisible even if the expansion is the same.

The Taylor expansion discussed in Merlini and Rusconi (2002), applied here to the potential $\varphi$ is given by:

$$\varphi(\rho) = \exp[-(\rho_o-\rho)\cdot\partial\rho_o]\varphi(\rho_o) = \sum (-1)^n (1/n!)\cdot(\rho_o-\rho)^n\cdot\partial\rho_o^n\ \varphi(\rho_o)$$

where n runs from 0 to $\infty$.

Here we are not interested in the value $\rho_o = 1^+$ outside the critical strip but in the value $\rho_o = 1^-$, which is in the critical strip, where we still have an expression for $\varphi(\rho)$ which we have proven; the Taylor expansion around $\rho_o=1-$ exists, has a convergent radius of ½, and the information on $\varphi(\rho)$ inside the strip is offered by the complete knowledge of the function around $\rho_o=1^-$ inside the critical strip. For $\rho_o = 1^-$ and $\rho = ½$, the expansion is given by

$$\varphi(1/2) = \lim_{n\to\infty}(\prod_{k=1}^{n}[1-1/k)\cdot(\rho-1/2)\partial\rho_0])\varphi(\rho_0)|\rho_0 = 1$$

where $\varphi(\rho_o) = \pi\cdot\ln(Zeta(\rho_o+½)\cdot(\rho_o-½)/(3/2-\rho_o))$, at $\rho_o = 1^-$.

Finally, if we assume the RH is true, then from Fig.1 it is seen that $\varphi(\rho)$ is not injective (the same for S(x) of Fig.4) and it may be of some interest in finding the solution of the equation

$$\varphi(\rho) = \pi\cdot\ln[(Zeta(\rho+½)\cdot(\rho-½)/(3/2-\rho)] = \varphi(\rho') = \pi\cdot\ln[Zeta(\rho'+½)] \qquad (19)$$

This implies a relation between two different value $\rho$ and $\rho'$ which give the same value of the potential $\varphi$.
Near $\rho = ½$ ($\rho' = \infty$) we have, using the crude approximation for the Zeta Function $\zeta(\rho') \sim 1+1/2^{\rho'}$ at $\rho' \sim \infty$ that:

$$\pi(1+\gamma)\cdot(\rho-½) \sim \pi\cdot(1/2^{\rho'})$$

and
$$\rho' = -\ln[(1+\gamma)\cdot(\rho-½)]/\ln 2.$$

Thus

$$\varphi(\rho) = \varphi(-\ln[(1+\gamma)\cdot(\rho-½)]/\ln 2)$$

near $\rho= ½$, if the RH is true. The values of $\varphi(\rho)$ are located in the range of the values of $\rho+½$ given by the interval [1,1.001]. On the other hand, without the above approximation for $\zeta(\rho'+½)$, $\rho(\rho') \sim 1/2 + \ln(Zeta(\rho'+½)/(1+\gamma)$, (still using the first order expansion for $\varphi(\rho)$ near $\rho= ½$) and we have asymptotically as

$$\varphi(\rho(\rho')) = \varphi(\rho') = \pi\cdot\ln(Zeta(\rho'+½)).$$

Here the values of $\varphi(\rho)$ are also good for big values of $\rho'$ up to about the value of $\rho' = 3$, starting at $\rho'= \infty$.

It would be interesting to know more about the meaning of the above relation (Eq(19)) i.e. $\varphi(\rho) = \varphi(\rho'(\rho))$ to search as a kind of "hidden symmetry" ( Merlini, 2003 ) and to see whether it has a concrete relation with the semi-positive property of the Li's coefficients which ensures the truth of the RH, Li (1997) ,see also Smith (1998). If we use the function of Fig.4 only, we observe that such a "symmetry" is invisible ; on the other hand, such a "symmetry" is visible on our plots (Fig.(1) and Fig(4)). A relation in the finite interval $[1, x_1]$ of x for the potential well $S(x)$ is also suggested by the plot of Fig.4 ; in a future work we will investigate others similar functions of the variable $t = Im(s)$.

**Acknowledgements**

It is a pleasure to thank Prof.Dr.Sergio Albeverio for reading the manuscript and for offering us very important remarks and helps for the improvement of the form of the manuscript.

**Footnote**

In the first draft of the manuscript we used the spelling "magnetization" m instead of that of potential $\varphi$. The spelling "phase transition" is also now replaced by "jump" in the electrical field. (It should be said that if one considers the usual map $z = 1-1/s$, where $s = \rho + i \cdot t$, the image of the pole $\rho = 1$ is $z = 0$ and $z=1$ is the accumulation point of the image of the critical line (given by the unit circle) for large values of t on it as well as of the images of the trivial zeros at big values of the integers n, given by $z = (1 + 1/(2 \cdot n))$. If the Riemann Zeta function is seen as a complex "partition function", there is a double accumulation of zeros at $z =1$ and following the picture of statistical mechanics this could be seen as a "phase transition"; at $z = 1$, big values $t_o$ of the imaginary part of the non trivial zeros "meet" big values of the integers n.